\documentclass[epjCONF]{svjour}

\usepackage{amssymb,amsmath}
\usepackage{graphicx}
\usepackage[varg]{txfonts} 
\usepackage[latin1]{inputenc}
\usepackage{algorithm,algorithmic}

\session-title{First International Conference on Numerical Physics}

\begin{document}

\title{Rare event sampling with stochastic growth algorithms}

\author{Thomas Prellberg\inst{1}\fnmsep\thanks{\email{t.prellberg@qmul.ac.uk}}}
\institute{School of Mathematical Sciences, Queen Mary University of London\\ 
Mile End Road, London E1 4NS, United Kingdom}

\abstract{
We discuss uniform sampling algorithms that are based on stochastic growth methods, using sampling of extreme configurations of polymers in simple lattice models as a motivation.
We shall show how a series of clever enhancements to a fifty-odd year old algorithm, the Rosenbluth method, led to a cutting-edge algorithm capable of uniform sampling of equilibrium statistical mechanical systems of polymers in situations where competing algorithms failed to perform well. Examples range from collapsed homo-polymers near sticky surfaces to models of protein folding.
}

\maketitle

\section{Introduction}

A large class of sampling algorithms are based on Markov Chain Monte Carlo methods \cite{landau2005}.
Here we shall introduce an alternative method of sampling based on stochastic
growth methods. Stochastic growth means that one attempts to randomly grow configurations of interest from 
scratch by successively increasing the system size (usually up to a desired maximal size). 

For simplicity, we shall restrict ourselves to the setting of lattice path models of linear polymers,
that is, models based on random walks configurations on a regular lattice, such as the square or the simple
cubic lattice. If we impose self-avoidance, i.e. if we forbid those random walk configurations that repeatedly
visit the same lattice site, we obtain the model of Self-avoiding Walks (SAW), used to describe polymers in a
good solvent. Monte-Carlo Simulations of SAW have been proposed as early as 1951 \cite{king1951}. Extensions of 
SAW have been used to study a variety of different phenomena, such as polymer collapse, adsorption of polymers 
at a surface, and protein folding. 

While this setting is rich enough to allow for the simulation of physically relevant scenarios, it is also
simple enough to serve as the ideal background for the description of the particular class of stochastic
growth algorithms which we shall describe here.

In Section \ref{rm} we briefly discuss simple sampling of SAW, review Rosenbluth sampling as the basic algorithm, and by combining this with  pruning and enrichment strategies, discuss the Pruned and Enriched Rosenbluth Method (PERM), and its extension to 
uniform sampling, flatPERM. In Section \ref{ext} we conclude with a description of an extension of stochastic 
growth methods to settings beyond linear polymers, called Generalized Atmospheric Rosenbluth Method (GARM).

\section{Sampling of Self-Avoiding Walks}
\label{rm}

\begin{figure}[ht]
\begin{center}
\includegraphics[width=0.65\textwidth]{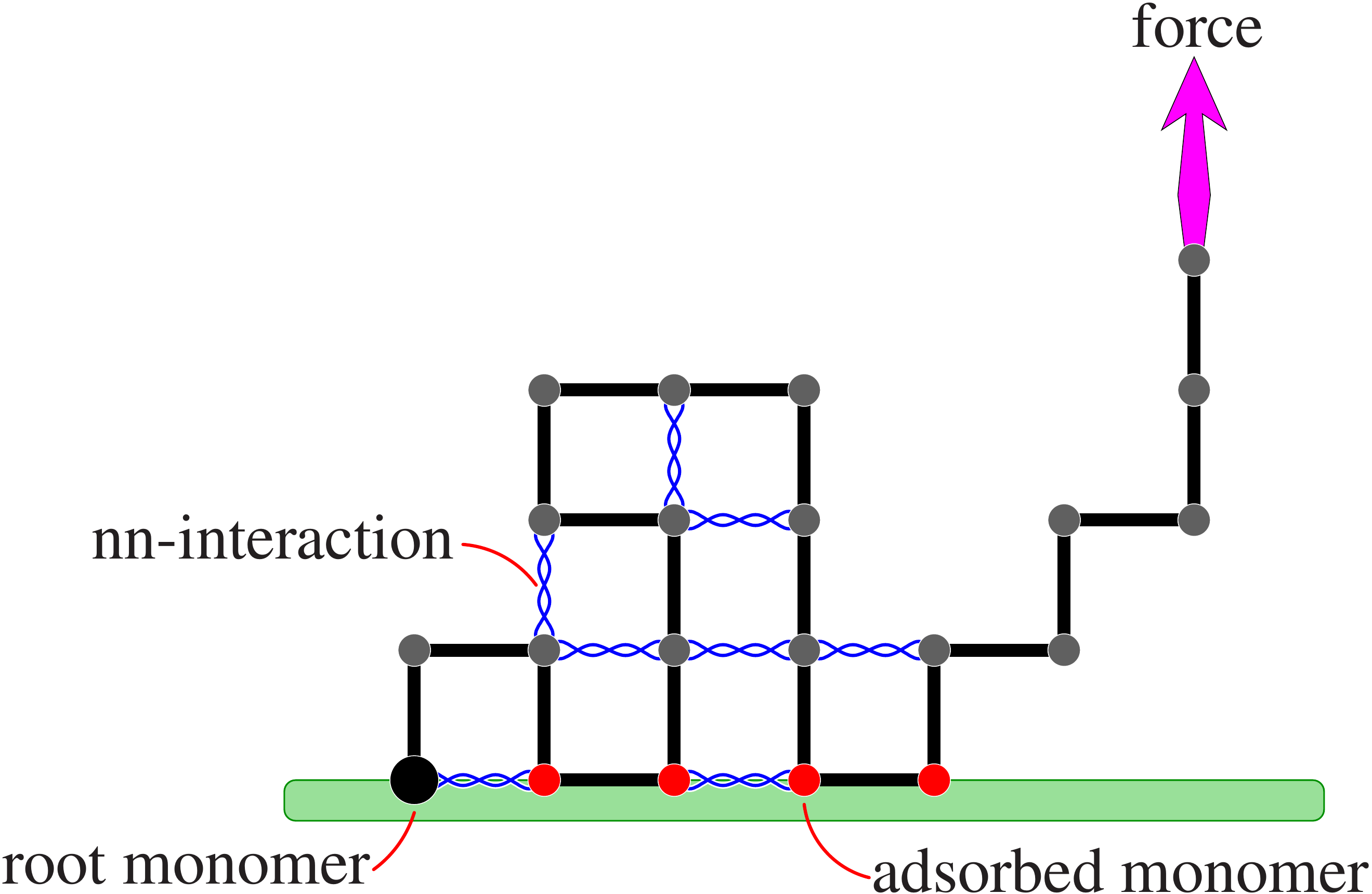}
\end{center}
\caption{A lattice model of a polymer tethered to a sticky surface under the influence of a pulling force.}
\label{polymer}
\end{figure}

In this section we want to consider the simulation of self-avoiding random walks (SAW), i.e. random 
walks obtained by forbidding any random walk that contains multiple visits to a lattice site. SAW is the 
canonical lattice model for polymers in a good solvent.  Moreover, it forms 
the basis for more realistic models of polymers with physically and biologically relevant structure, as
indicated in Figure \ref{polymer}.

However, the introduction of self-avoidance turns a simple Markovian random walk without memory into a 
complicated non-Markovian random walk; when growing a self-avoiding walk, one needs to test for 
self-intersection with all previous steps, leading to a random walk with infinite memory.

\subsection{Simple Sampling}
\label{simplesaw}

\begin{algorithm}[ht]
\caption{Simple Sampling of Self-Avoiding Walk}
\label{sssaw}
\begin{algorithmic}
   \STATE $s_{\cdot}\gets0$
   \STATE $Samples\gets0$
   \WHILE {$Samples<MaxSamples$}
      \STATE $Samples\gets Samples+1$
      \STATE $n\gets0$, Start at origin
      \STATE $s_{0}\gets s_{0}+1$
      \WHILE {$n<MaxLength$}
         \STATE Draw one of the neigboring sites uniformly at random
         \IF {Occupied}
             \STATE Reject entire walk and exit loop
         \ELSE
             \STATE Step to new site
             \STATE $n\gets n+1$
             \STATE $s_{n}\gets s_{n}+1$
         \ENDIF
      \ENDWHILE
   \ENDWHILE
\end{algorithmic}
\end{algorithm}

It is straight-forward to generate SAW by simple sampling. Generating an $n$-step self-avoiding walk with
the correct statistics, i.e. such that every walk is generated with the same probability, is equivalent to generating
$n$-step random walks and reject those random walks that self-intersect. Algorithm \ref{sssaw} accomplishes this
by generating two-dimensional random walks and rejecting the \emph{complete} configuration when self-intersection
occurs.

At each step, the walk has four possibilities to continue, and chooses one of these with probability $p=1/4$. Therefore
an estimator for the total number of $n$-step SAW after $S$ samples have been generated is given by $4^ns_n/S$.

\begin{figure}[ht]
\begin{center}
\includegraphics[width=0.75\textwidth]{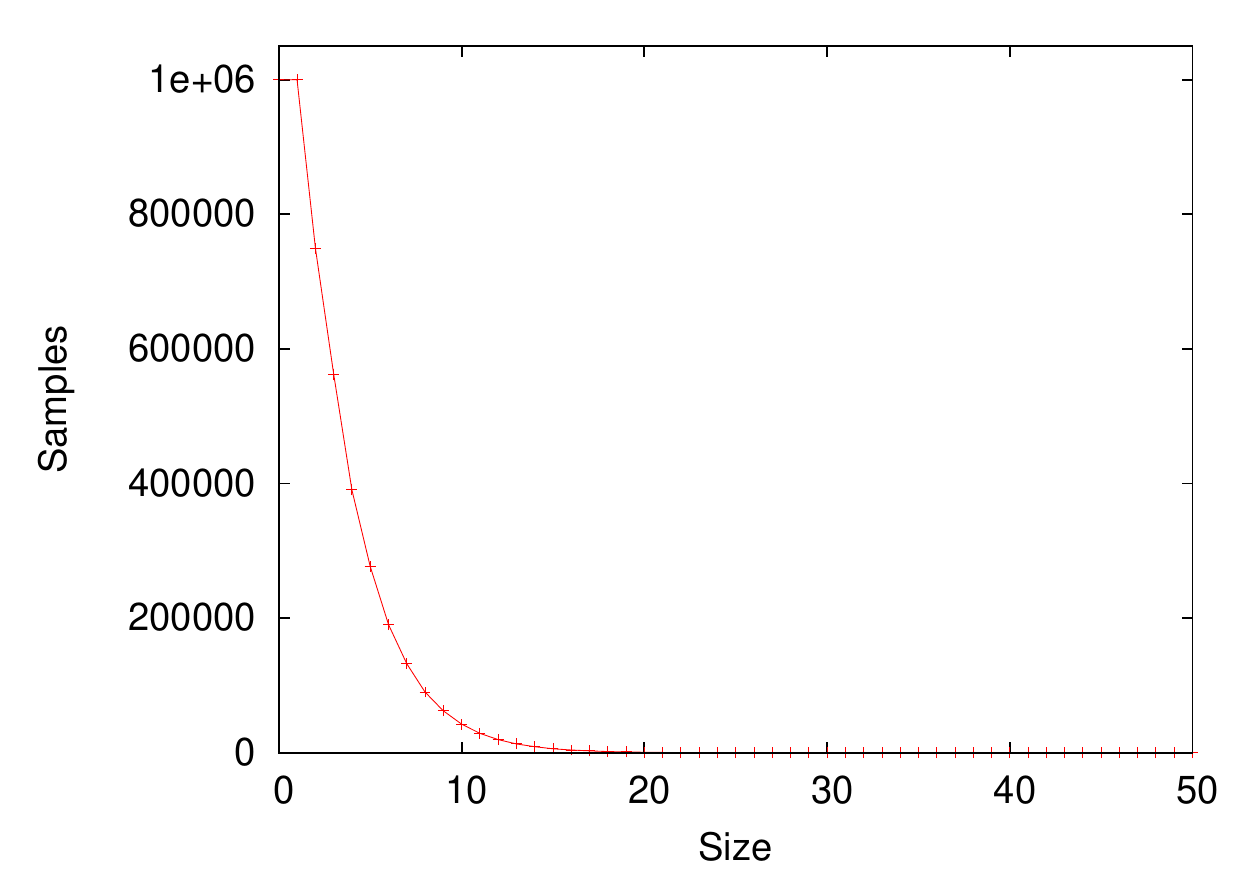}
\end{center}
\caption{Attrition of started walks generated with Simple Sampling. From $10^6$ started walks none grew more than 35 steps.}
\label{figure5}
\end{figure}

Generating SAW with simple sampling is very inefficient. There are $4^n$ $n$-step random walks, but only about $2.638^n$
$n$-step self-avoiding walks on the square lattice. The probability of successfully generating an $n$-step self-avoiding
walk therefore decreases exponentially fast, leading to very high attrition\footnote{The algorithm can be improved somewhat by forbidding immediate reversals of the random walk, but the attrition remains exponential}. Longer walks 
are practically inaccessible, as seen in Figure \ref{figure5}.

\subsection{Rosenbluth Sampling}
\label{rosenbluthsaw}

A slightly improved sampling algorithm was proposed in 1955 by Rosenbluth and Rosenbluth \cite{rosenbluth1955}. The basic idea is to avoid
self-intersections by only sampling from the steps that lead to self-avoiding configurations. In this way, the algorithm
only terminates if the walk is trapped in a dead end and cannot continue growing. While this still happens exponentially
often, Rosenbluth sampling can produce substantially longer configurations than simple sampling.

While simple sampling generates all configurations with equal probability, configurations generated with 
Rosenbluth sampling are generated with different probabilities. To understand this in detail, it is 
helpful to introduce the notion of an \emph{atmosphere} of a configuration; this is the number of ways 
in which a configuration can continue to grow. For one-dimensional simple random walks the atmosphere is always two, for
two-dimensional simple random walks on the square lattice the atmosphere is always four (and if one forbids immediate self-reversals, the atmosphere is always three except for the very first step). However, for self-avoiding walks on 
the square lattice the atmosphere is a configuration-dependent quantity assuming values between four (for the first step)
and zero (for a trapped configuration that cannot be continued). We shall denote the atmosphere of a configuration $\phi$
by $a(\phi)$. If it is clear from the context, we will drop the argument and speak about the atmosphere $a$.

If a configuration has atmosphere $a$, this means that there are $a$ different possibilities of growing the configuration,
and each of these can get selected with probability $p=1/a$. To balance this, the weight of this configuration is
therefore multiplied by the atmosphere $a$. An $n$-step walk grown by Rosenbluth sampling therefore has weight
$$W_n=\prod_{i=0}^{n-1}a_i\;,$$
where $a_i$ are the atmospheres of the configuration after $i$ growth steps. This walk is generated with probability 
$P_n=1/W_n$, so that $P_nW_n=1$ as required. Algorithm \ref{rsaw} shows a pseudocode implementation of Rosenbluth sampling.
\begin{algorithm}[ht]
\caption{Rosenbluth Sampling of Self-Avoiding Walk}
\label{rsaw}
\begin{algorithmic}
   \STATE $s_{\cdot}\gets0$, $w_{\cdot}\gets0$
   \STATE $Samples\gets0$
   \WHILE {$Samples<MaxSamples$}
      \STATE $Samples\gets Samples+1$
      \STATE $n\gets0$, $Weight\gets 1$, Start at origin
      \STATE $s_{0}\gets s_{0}+1$, $w_0\gets w_0+Weight$
      \WHILE {$n<MaxLength$}
         \STATE Create list of neighboring unoccupied sites, determine the atmosphere $a$
         \IF {$a=0$ (walk cannot continue)}
             \STATE Reject entire walk and exit loop
         \ELSE
             \STATE Draw one of the neigboring unoccupied sites uniformly at random
             \STATE Step to new site
             \STATE $n\gets n+1$, $Weight\gets Weight\times a$
             \STATE $s_{n}\gets s_{n}+1$, $w_n\gets w_n+Weight$
         \ENDIF
      \ENDWHILE
   \ENDWHILE
\end{algorithmic}
\end{algorithm}

\begin{figure}[ht]
\begin{center}
\includegraphics[width=0.75\textwidth]{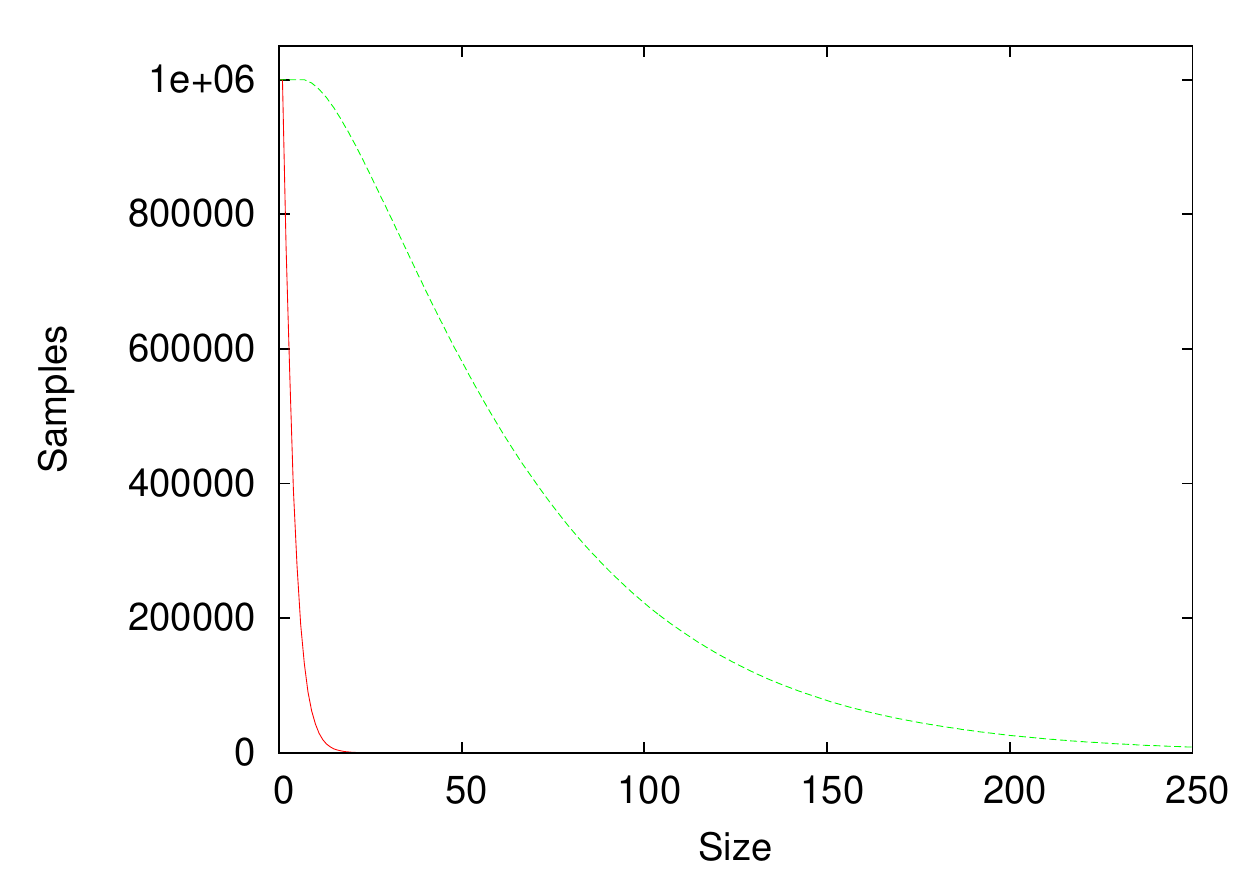}
\end{center}
\caption{Attrition of started walks generated with Rosenbluth Sampling compared with Simple Sampling. Walks with a few hundred steps become accessible.}
\label{figure6}
\end{figure}

Figure \ref{figure6} shows the improvement gained by Rosenbluth sampling over simple sampling.

\subsection{Pruned and Enriched Rosenbluth Sampling}
\label{perm}

It took four decades before Rosenbluth sampling was improved upon. In 1997 Grassberger augmented Rosenbluth
sampling with pruning and enrichment strategies, calling the new algorithm Pruned and Enriched Rosenbluth Method,
or PERM \cite{grassberger1997}. There are a variety of pruning and enrichment strategies that are possible, and
the strategies used in \cite{grassberger1997} were somewhat different from the ones we shall describe now. For alternate versions and enhancements we also refer to \cite{buks2009} and references therein. 

Suppose a walk has been generated that has weight $w$ as opposed to a target weight $W$. In the ideal situation $w$ is
equal to $W$ as desired. If that is not the case, either the weight $w$ is to small, i.e. the ratio $R=w/W<1$, or the
weight $w$ is too large, i.e. $R=w/W>1$. In the first case we will employ pruning, i.e. we will probabilistically remove
walks.
\begin{itemize}
\item If $R=w/W<1$, continue growing with probability $R$ and weight $w$ set to $W$, and stop growing with probability $1-R$.
\end{itemize}
In the second case we will employ enrichment, i.e. we will continue to grow multiple copies of the walk.
\begin{itemize}
\item If $R=w/W>1$, make $\lfloor R\rfloor+1$ copies with probability $p=R-\lfloor R\rfloor$ and $\lfloor R\rfloor$ copies with probability 
$1-p$. Continue growing with the weight of each copy set to $W$.
\end{itemize}
While we chose to describe pruning and enrichment as different strategies, note that enrichment procedure is actually identical
to the pruning procedure if $R<1$: when $\lfloor R\rfloor=0$ then enrichment reduces to making $1$ copy with probability $R$ and
$0$ copies with probability $1-R$, which is just the pruning procedure.

Whereas in simple (or Rosenbluth) sampling the generated walks are each grown independently from length zero, pruning and enrichment 
leads to the generation of a large tree-like structure of more or less correlated walks grown from one seed. We call the collection
of these walks a \emph{tour} of the algorithm. The tree structure of a tour allows for successively growing all copies obtained 
during the enrichment in a natural way. 

Pruning and enrichment can be incorporated quite easily as follows.

\begin{algorithmic}
\REPEAT
   \IF {zero atmosphere or maximal length reached}
      \STATE set number of enrichment copies to zero
   \ELSE
      \STATE prune/enrich step: compute number of enrichment copies
   \ENDIF
   \IF {number of enrichment copies is zero}
       \STATE prune: shrink to previous enrichment
   \ENDIF
   \IF {configuration shrunk to zero}
      \STATE start new tour
      \STATE store data for new configuration
   \ELSE
      \STATE decrease number of enrichment copies
      \IF {positive atmosphere}
         \STATE grow new step
         \STATE store data for new configuration
      \ENDIF
   \ENDIF
\UNTIL {enough data is generated}
\end{algorithmic}

Note that in case of constant atmosphere this reduced precisely to the pruned and enriched sampling for
simple random walks encountered earlier.
Algorithm \ref{permsaw} contains a more detailed pseudo-code version of PERM for self-avoiding walks.

\begin{algorithm}[ht]
\caption{Pruned and Enriched Rosenbluth Sampling of Self-Avoiding Walks}
\label{permsaw}
\begin{algorithmic}
   \STATE $s_{\cdot}\gets0$, $w_{\cdot}\gets0$
   \STATE $Tours\gets0$, $n\gets0$, $Weight_0\gets1$
   \STATE Start new walk with step size zero
   \STATE $a\gets0$, $Copy_0\gets1$
   \STATE $s_{0}\gets s_{0}+1$, $w_0\gets w_0+Weight$
   \WHILE[Main loop] {$Tours<MaxTours$}
       \IF[Maximal length reached or atmosphere zero: don't grow] {$n=MaxLength$ or $a=0$}
         \STATE $Copy_n\gets0$
      \ELSE[pruning/enrichment by comparing with target weight]
         \STATE $Ratio\gets Weight_n/w_n$
         \STATE $p\gets Ratio\mod 1$
         \STATE Draw random number $r\in[0,1]$
         \IF {$r<p$}
            \STATE $Copy_n\gets\lfloor Ratio\rfloor+1$
         \ELSE
            \STATE $Copy_n\gets\lfloor Ratio\rfloor$
         \ENDIF
         \STATE $Weight_n\gets w_n$
      \ENDIF         
      \IF[Shrink to last enrichment point or to size zero] {$Copy_n=0$}
         \WHILE {$n>0$ and $Copy_n=0$}
            \STATE Delete last site of walk
            \STATE $n\gets n-1$
         \ENDWHILE
      \ENDIF
      \IF[start new tour] {$n=0$ and $Copy_0=0$}
         \STATE $Tours\gets Tours+1$, 
         \STATE Start new walk with step size zero
         \STATE $a\gets0$, $Copy_0\gets1$
         \STATE $s_{0}\gets s_{0}+1$, $w_0\gets w_0+Weight$
      \ELSE
         \STATE Create list of neighboring unoccupied sites, determine the atmosphere $a$
         \IF {$a>0$}
            \STATE $Copy_n\gets Copy_n-1$
            \STATE Draw one of the neigboring unoccupied sites uniformly at random
            \STATE Step to new site
            \STATE $n\gets n+1$, $Weight_n\gets Weight_n\times a$
            \STATE $s_{n}\gets s_{n}+1$, $w_n\gets w_n+Weight_n$
         \ENDIF
      \ENDIF
   \ENDWHILE
\end{algorithmic}
\end{algorithm}

\begin{figure}[ht]
\begin{center}
\includegraphics[width=0.75\textwidth]{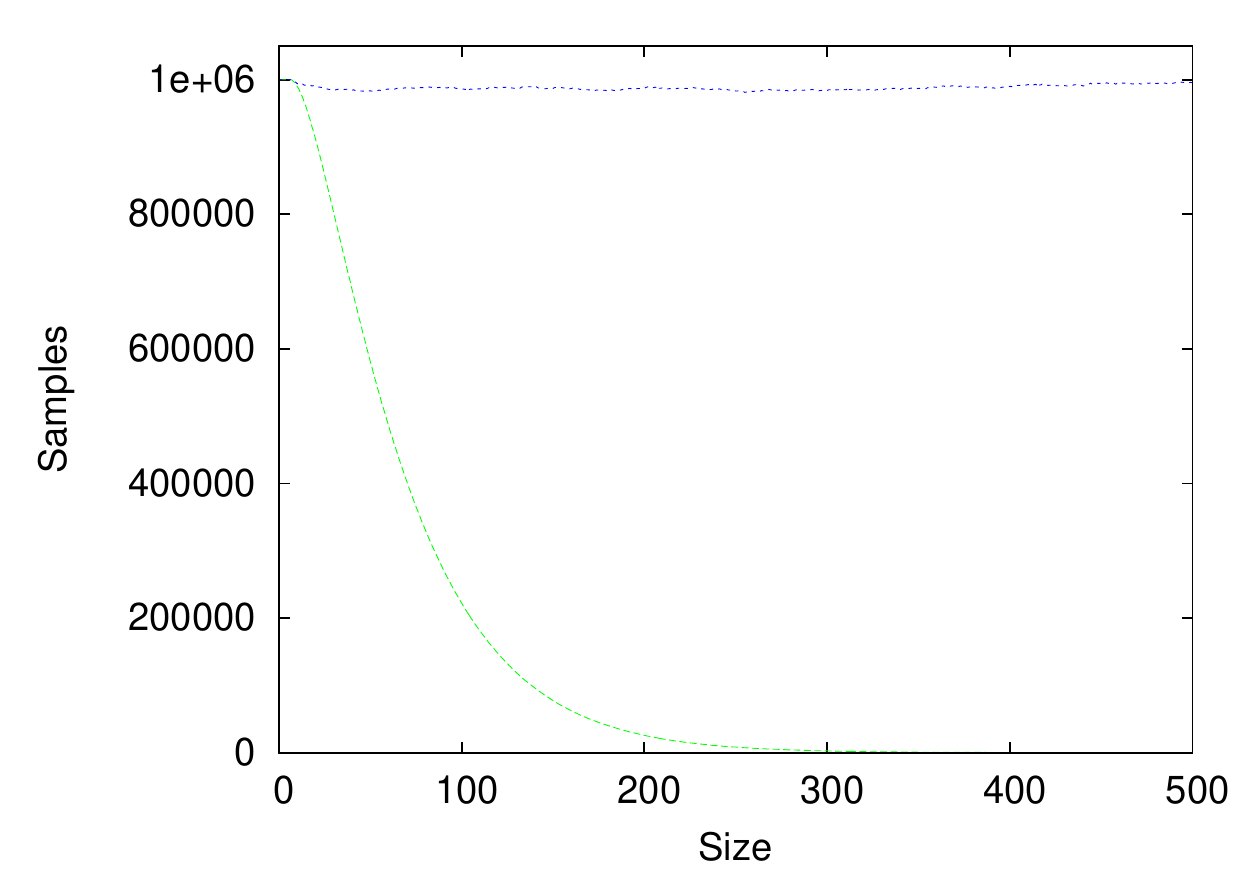}
\end{center}
\caption{Attrition of started walks with PERM compared with Rosenbluth Sampling. In the case of PERM, a virtually constant number of samples is obtained.}
\label{figure7}
\end{figure}

Figure \ref{figure7} shows the significant improvement gained by adding pruning and enrichment strategies to Rosenbluth
Sampling.

\subsection{Flat Histogram Rosenbluth Sampling}
\label{flatperm}

The next advance was made by two groups in 2003/4. Motivated by work of Wang and Landau on uniform sampling
\cite{wang2001}, Bachmann and Janke, using ideas from Berg and Neuhaus \cite{berg1991}, implemented 
Multicanonical PERM \cite{bachmann2003}. This was followed by Prellberg and Krawczyk \cite{prellberg2004}, who
designed flatPERM, a flat-histogram version of PERM estimating directly the microcanonical density of states.

Within the context of the algorithms developed here, incorporating uniform sampling into PERM is straightforward.
First we note that PERM already is a uniform sampling algorithm in system size. This is not apparent at all from the
algorithm, as the guiding principle has been to adjust pruning and enrichment with respect to a target weight, not
with respect to any criterion of poor local sampling. It is rather that uniform sampling is a consequence of adjusting
pruning and enrichment around the desired target weight.

\begin{figure}[ht]
\begin{center}
\includegraphics[width=0.5\textwidth]{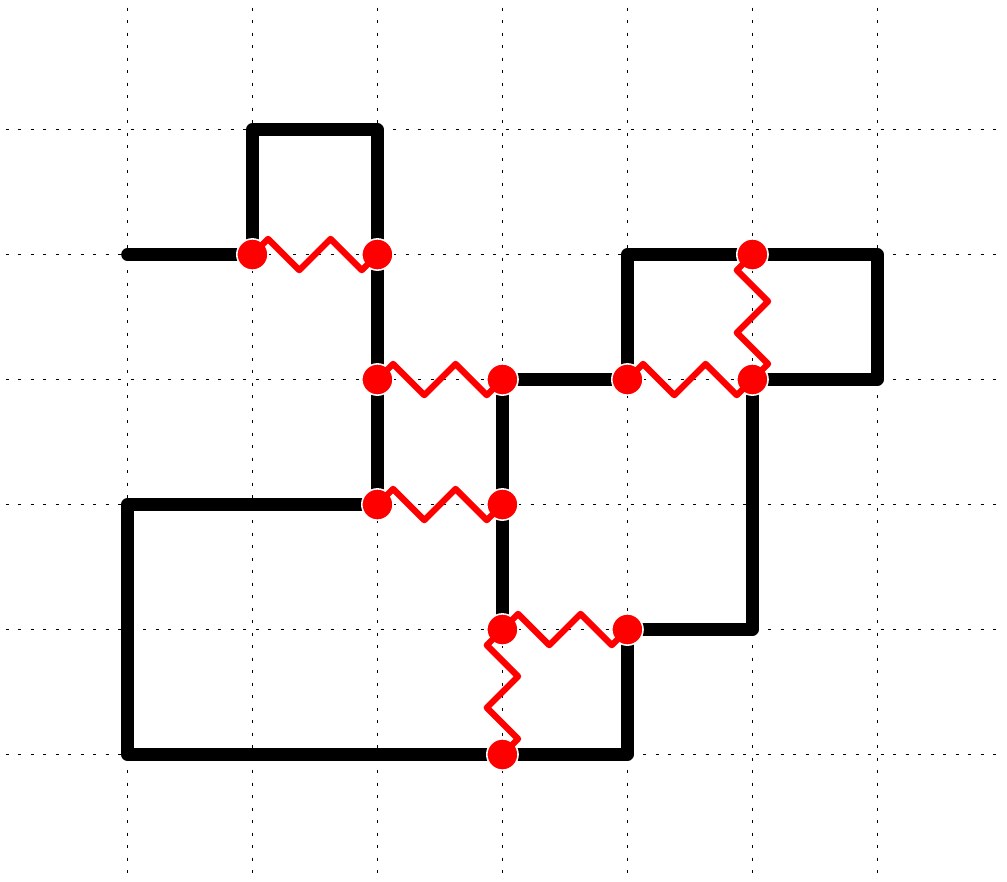}
\end{center}
\caption{An interacting self-avoiding walk on the square lattice with $n=26$ steps and $m=7$ contacts.}
\label{isaw}
\end{figure}

It is therefore reasonable (and very much in the spirit of the previous section) to extend PERM to a microcanonical
version, in which configurations of size $n$ are separated with respect some additional parameter. One simply
determines this parameter when growing the configuration and stores the data by binning with respect to this
additional parameter. Then, when considering pruning and enrichment, the target weight is computed from the
binned data. More precisely, if the additional parameter is called $m$, storing the data is changed from

\begin{algorithmic}
   \STATE $s_{n}\gets s_{n}+1$, $w_n\gets w_n+Weight_n$
\end{algorithmic}

to 

\begin{algorithmic}
   \STATE $s_{n,m}\gets s_{n,m}+1$, $w_{n,m}\gets w_{n,m}+Weight_n$
\end{algorithmic}

and computing enrichment ratio is changed from

\begin{algorithmic}
   \STATE $Ratio\gets Weight_n/w_n$
\end{algorithmic}

to 

\begin{algorithmic}
   \STATE $Ratio\gets Weight_n/w_{n,m}$
\end{algorithmic}

and this is about it.

In the previous section this additional parameter has been the end-point position of the random walk. Here, we 
shall consider by example the case of \emph{interacting self-avoiding walks}, where each walk configuration has 
an energy proportional to the number of non-consecutive nearest-neigbour contacts between occupied lattice sites.

\begin{figure}[ht]
\begin{center}
\includegraphics[width=0.75\textwidth]{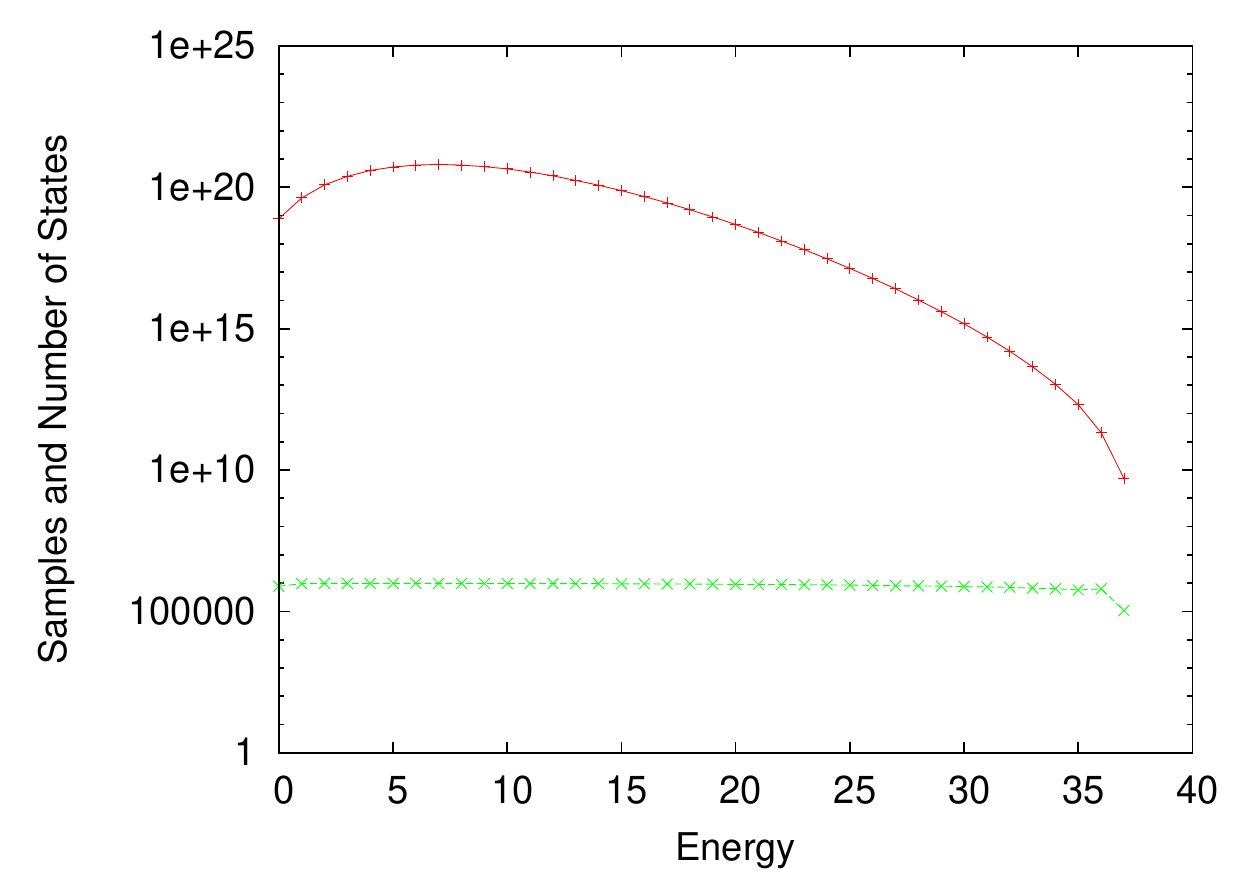}
\end{center}
\caption{Number of States of Interacting Self-Avoiding Walks with 50 steps at fixed energy estimated from $10^6$ 
flatPERM tours. The lower graph shows the number of actually generated samples for each energy.}
\label{figure9}
\end{figure}

Figure \ref{figure9} shows the simulation results of a simulation of interacting self-avoiding walks of up to $50$ 
steps using $10^6$ tours starting at size zero. This led to the generation of about $10^6$ samples for each value of
$m$ at $n=50$ steps, and enabled the estimation of the number of states over ten orders of magnitude.

\begin{figure}[ht]
\begin{center}
\includegraphics[width=0.75\textwidth]{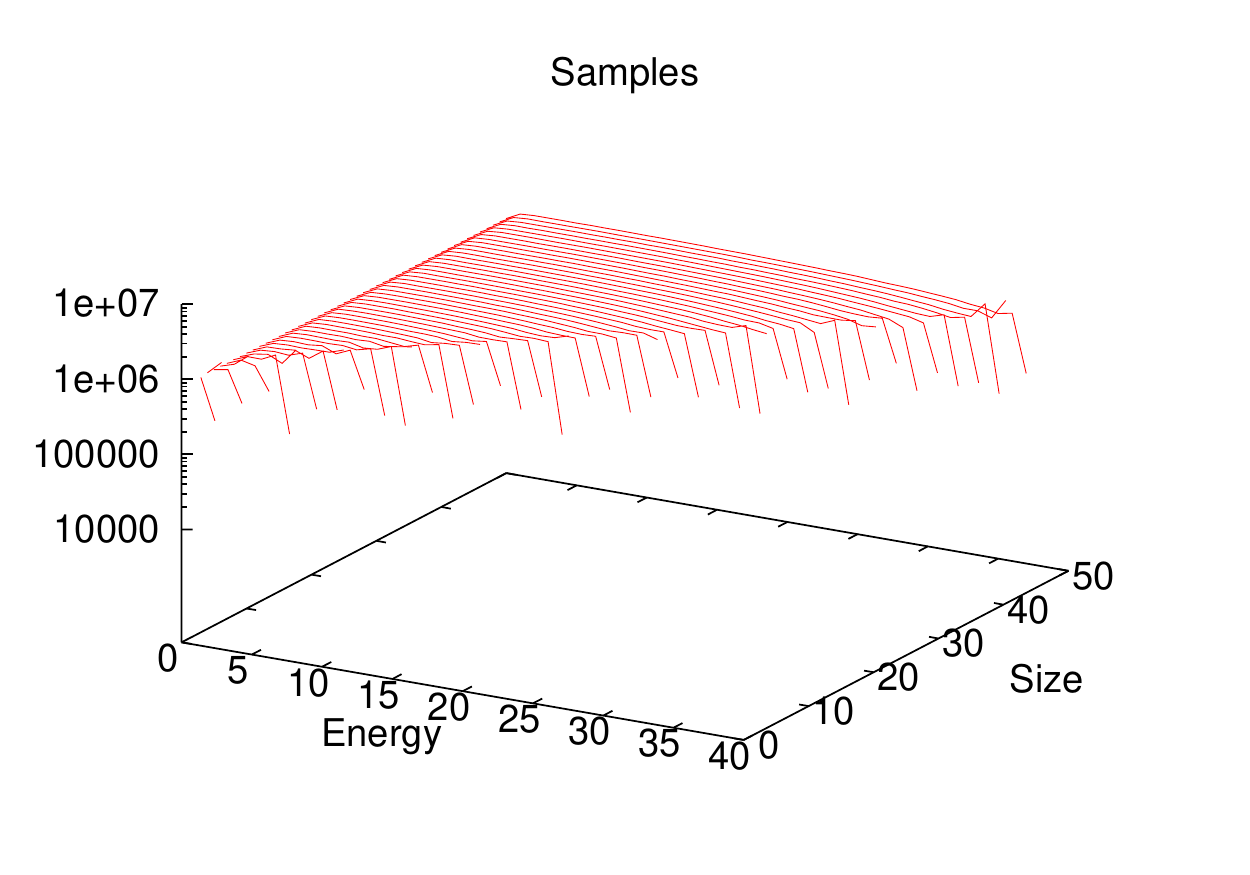}

\includegraphics[width=0.75\textwidth]{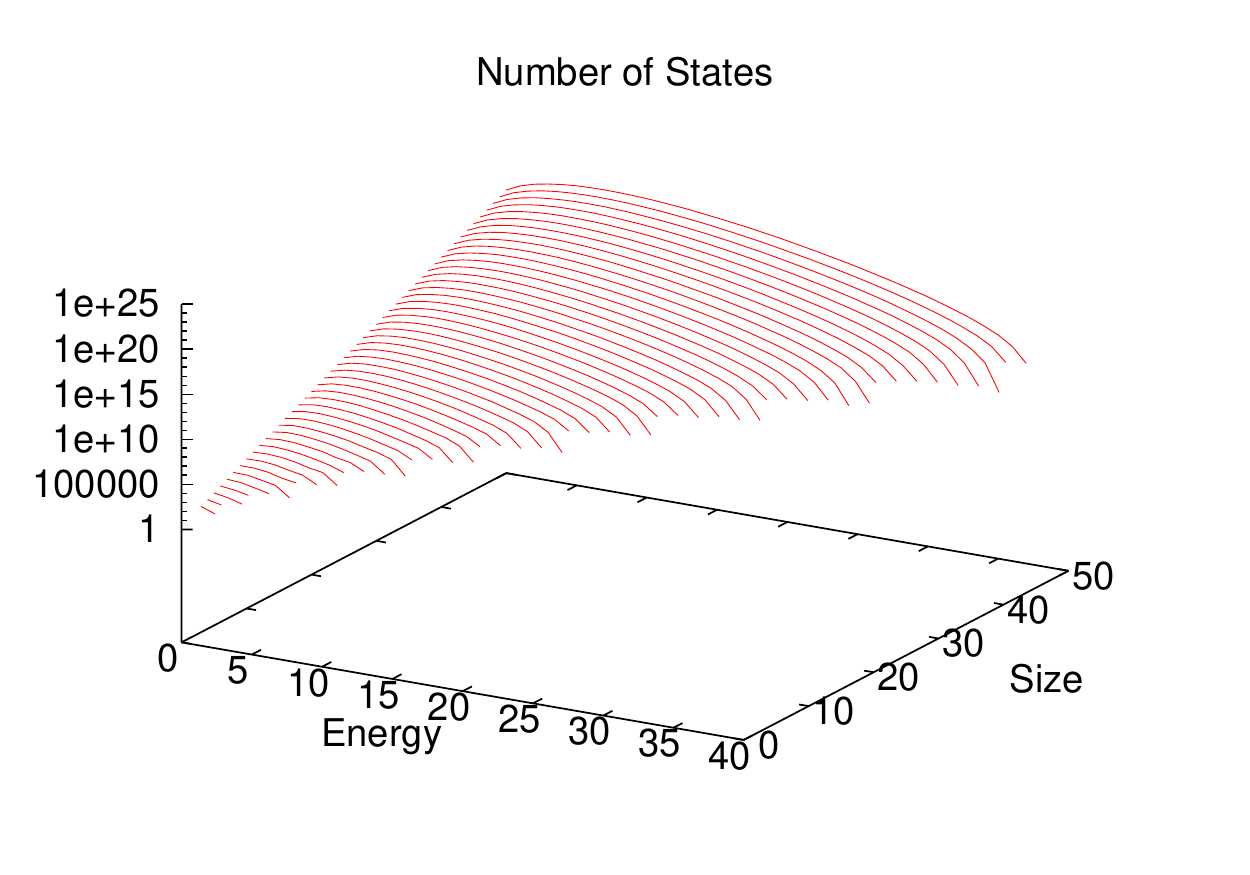}
\end{center}
\caption{Interacting Self-Avoiding Walks with up to 50 steps generated with flatPERM. The upper figure
shows that a roughly constant number of samples is obtained across the whole range of sizes and energies,
and the lower figure shows the estimated number of states for a given Size $n$ and Energy $m$.}
\label{figure8}
\end{figure}

Figure \ref{figure8} shows the corresponding simulation results for all intermediate lengths from the same run. While
the histogram of samples is not as flat as in the case of random walks discussed above, especially for large values of
$m$, it is reasonably flat on a logarithmic scale and leads to sufficiently many samples for each histogram bin.
Reference \cite{prellberg2004} contains results of a simulation of interacting self-avoiding walks with up to $n=1024$
steps, where the density of states ranges over three hundred orders of magnitude, all obtained from one single
simulation.

\section{Extensions}
\label{ext}

In the excellent review article \cite{buks2009} the algorithms of Section \ref{rm} and several extensions
are discussed. 

Extensions of algorithms are usually motivated by the need to simulate systems inaccessible with
established algorithm. For example, algorithms based on Rosenbluth sampling are well-suited to the simulation of 
objects that
can be grown uniquely from a seed. In the case of linear polymer models, this is accomplished by appending a step
to the end of the current configuration\footnote{In a more abstract setting, Rosenbluth sampling has for example
been used to study the number of so-called pattern-avoiding permutations. Permutations can be grown easily by inserting the number $n+1$
somewhere into a permutation of the numbers $\{1,2,\ldots,n\}$, allowing for easy implementation of Rosenbluth sampling.}.

However, if one wants to simulate polymers with a more complicated structure, such as branched polymers, there no 
longer is an easy way to uniquely grow a configuration. A lattice model for a two-dimensional branched polymer is 
given by lattice trees, i.e. trees embedded in the lattice $\mathbb Z^2$. For a given lattice tree it is no longer 
clear how it has been grown from a seed; this could have happened in a variety of ways.

\subsection{Generalized Atmospheric Rosenbluth Sampling}

It turns out that there is an extension to Rosenbluth sampling, called Generalized Atmospheric Rosenbluth Method, 
or GARM \cite{rechnitzer2008}, that is suitable for these more complicated growth processes. The key idea is to generalize the notion of atmosphere by introducing an additional negative
atmosphere $a^-$ indicating in how many ways a configuration can be reduced in size. For linear polymers the 
negative atmosphere is always unity, as there is only one way to remove a step from the end of the walk. However, 
for a given lattice tree the removal of any leaf of the tree gives a smaller lattice tree, and the negative 
atmosphere $a^-$ can assume rather large values.

\begin{algorithm}[ht]
\caption{Generalised Atmospheric Sampling}
\label{garm}
\begin{algorithmic}
   \STATE $s_{\cdot}\gets0$, $w_{\cdot}\gets0$
   \STATE $Samples\gets0$
   \WHILE {$Samples<MaxSamples$}
      \STATE $Samples\gets Samples+1$
      \STATE $n\gets0$, $Weight\gets 1$, Start with seed configuration
      \STATE $s_{0}\gets s_{0}+1$, $w_0\gets w_0+Weight$
      \WHILE {$n<MaxSize$}
         \STATE Create list of growth possibilities, determine the atmosphere $a$
         \IF {$a=0$ (no growth possible)}
             \STATE Reject entire configuration and exit loop
         \ELSE
             \STATE Draw one of the growth possibilities uniformly at random
             \STATE Grow configuration
             \STATE $n\gets n+1$, $Weight\gets Weight\times a$
             \STATE Compute negative atmosphere $a^-$
             \STATE $Weight\gets Weight/a^-$
             \STATE $s_{n}\gets s_{n}+1$, $w_n\gets w_n+Weight$
         \ENDIF
      \ENDWHILE
   \ENDWHILE
\end{algorithmic}
\end{algorithm}

Surprisingly there is a very simple extension to the Rosenbluth weights discussed above.
If a configuration has negative atmosphere $a^-$, this means that there are $a^-$ different possibilities in which the
configuration could have been grown. An $n$-step configuration grown by GARM therefore has weight 
\begin{equation}
W_n=\prod_{i=0}^{n-1}\frac{a_i}{a_{i+1}^-}\;,
\end{equation}
where $a_i$ are the (positive) atmospheres of the configuration after $i$ growth steps, and $a_i^{-}$ are the negative atmospheres of the configuration after $i$ growth steps. It can be shown that the probability of growing this configuration is
$P_n=1/W_n$, so again $P_nW_n=1$ holds as required.

The implementation of GARM is not any more complicated than the implementation of Rosenbluth sampling. Algorithm
\ref{rsaw} gets changed minimally by inserting the lines
\begin{algorithmic}
\STATE Compute negative atmosphere $a^-$
\STATE $Weight\gets Weight/a^-$
\end{algorithmic}
immediately after having grown the configuration.

While implementing GARM is quite straightforward, there generally is a need for more complicated data structures 
for the simulated objects, and one needs to find efficient algorithms for the computation of positive and negative atmospheres.

It is now possible to add pruning and enrichment to GARM, and to extend this further to flat histogram sampling,
just as has been described in the previous section for Rosenbluth sampling.

For further extensions to Rosenbluth sampling, and indeed many more algorithms for simulating self-avoiding 
walks, as well as applications, see \cite{buks2009}.


\end{document}